\documentclass[pra,twocolumn,superscriptaddress,floatfix,amssymb,showpacs
]{revtex4-1}

\usepackage{graphicx}% Include figure files
\usepackage{dcolumn}% Align table columns on decimal point
\usepackage{bm}% bold math
\usepackage[fleqn]{amsmath}
\newcommand{\RN}[1]{%
  \textup{\uppercase\expandafter{\romannumeral#1}}%
}
\usepackage{lipsum}
\usepackage{mwe}
\usepackage{hyperref}
\usepackage{hypernat}
\begin{document}

%\preprint{APS/123-QED}
\title{Room-Temperature Photon-Number-Resolved Detection Using A Two-Mode Squeezer}

%\thanks{A footnote to the article title}%

\author{Elisha S. Matekole}
\email{esiddi1@lsu.edu}
\affiliation{Hearne Institute for Theoretical Physics and Department of Physics and Astronomy, Louisiana State University,
Baton Rouge, Louisiana 70803, USA.}

\author{Deepti Vaidyanathan}
\affiliation{Baton Rouge Magnet High School, Louisiana 70806, USA.}

\author{Kenji W. Arai}
\affiliation{Reed College, Portland, Oregon 97202-8199, USA.}

\author{Ryan T. Glasser}
\affiliation{Department of Physics and Engineering Physics, Tulane University, New Orleans, LA 70118, USA.}  

\author{Hwang Lee}
\affiliation{Hearne Institute for Theoretical Physics and Department of Physics and Astronomy, Louisiana State University,
Baton Rouge, Louisiana 70803, USA.} 

\author{Jonathan P. Dowling} 
\affiliation{Hearne Institute for Theoretical Physics and Department of Physics and Astronomy, Louisiana State University,
Baton Rouge, Louisiana 70803, USA.} 
\date{\today}% It is always \today, today,
             %  but any date may be explicitly specified

%\begin{abstract}
%An article usually includes an abstract, a concise summary of the work
%covered at length in the main body of the article. 
%\begin{description}
%\item[Usage]
%Secondary publications and information retrieval purposes.
%\item[PACS numbers]
%May be entered using the \verb+\pacs{#1}+ command.
%\item[Structure]
%You may use the \texttt{description} environment to structure your abstract;
%use the optional argument of the \verb+\item+ command to give the category of each item. 
%\end{description}
%\end{abstract}

\begin{abstract}
\noindent{We study the average intensity-intensity correlations signal at the output of a two-mode squeezing device with $|N\rangle\otimes|\alpha\rangle$ as the two input modes. We show that the input photon-number can be resolved from the average intensity-intensity correlations. In particular, we show jumps in the average intensity-intensity correlations signal as a function of input photon-number $N$. Therefore, we propose that such a device may be deployed as photon-number-resolving detector at room temperature with high efficiency.}

%%Valid PACS numbers may be entered using the \verb+\pacs{#1}+ command.
\end{abstract}
  %  Quantum fluctuations, quantum noise, and quantum jumps

%Use showkeys class option if keyword
                              %display desired
\maketitle

%\tableofcontents

\section{\label{sec:level1} INTRODUCTION}
Photon-number-resolving detectors (PNRD) are crucial to the field of quantum optics, and quantum information processing. 
%including to quantum repeaters, and for linear optics quantum computing
%\cite{simon2007, knill2001, dowling2007}. 
PNRD can be useful in two major classes of application:Single-shot measurement of photon number, and ensemble measurements for photon number statistics. Single-shot photon number measurement is useful in the field of linear optical quantum computing, quantum repeaters, entanglement swapping, and conditional state preparation 
\cite{knill2001,dowling2007,simon2007,Tittel2009,Sliwa2003}.
%entanglement swapping). 
%PNRD is useful in the operation of entanglement swapping gate, which requires post-selection based on one and two photons.
%\cite{obrien2007}.

Ensemble measurement based PNRD can be used in quantum interferometry for measuring photon statistics, characterization of quantum light sources, and improvement in sensitivity and resolution.
\cite{shields2007,Lincoln2000, APD,wild2009,Vogel2016}. 
For example, a true single-photon source is important for quantum key distribution. The ultimate security of the key can be compromised if the source emits more than one photon in the same quantum bit state. Hence, a photon-number resolving detector that can characterize the single-photon source accurately is vital for the success of quantum key distribution
\cite{Sanders2000,Miller2003}. 
%cite QKD.
Also, the reconstruction of photon-statistics of unknown light sources by ensemble measurements can be used to determine the nature of the light source (classical or non-classical), and detection of weak thermal light, and coherent light.
Therefore, a desirable feature of a PNRD is accurate detection of the number of photons. In this paper, we propose a room-temperature photon-number-resolving detector using a two-mode squeezing device that finds its application in the reconstruction of photon statistics of unknown light states, and characterization of non classical light resources. For example, source characterization for enhanced quantum key distribution, and detection of weak thermal light. 

Commonly used photon detectors are the bucket or on/off detectors. These detectors can only distinguish between zero or more photons. 
Photon-number-resolving detectors typically include avalanche-based photodiodes, such as the visible light photon counters
\cite{APD, APD1}, 
two-dimensional arrays of avalanche photodiodes
\cite{yama2006, jiang2007},
time-multiplexed detectors
\cite{chilli2003, fitch2003,chilli2006},
photomultipliers
\cite{Zambra2004},
and weak avalanche-based PNRD
\cite{Kardynal2008}.
Most of these detectors have a high dark-count rate at room temperature, and are not sensitive to photon number greater than one. Therefore, they cannot be used in applications that require photon statistics. 
Another type of PNRD is a transition edge sensor (TES), which is a superconducting microbolometer.
These detectors are highly efficient but they operate at extremely low temperatures and have a low response time 
%\cite{Miller2003}, 
\cite{Miller2003,Lita2008a,Lita2008b,Calkins2013}.
Another superconductor-based PNRD uses parallel superconducting-nanowires, which can resolve finite number of photons at telecommunication wavelengths
\cite{Divochiy2008,Marsili2009}. 
Recently, atomic-vapor-based photon-number-resolving detectors have also been proposed
\cite{Zhifan2016}.
The merit of any PNRD is determined by detector efficiency, dark count rate, and response time. Most of the current photon-number-resolving detectors either have low efficiency or are plagued with high dark-count rates and low response time. Moreover, they have to be maintained at extremely low temperature to yield high efficiency. 

A two-mode squeezed vacuum (TMSV), also known as the twin-beam state, is an entangled state containing strong correlations between the two beams. However, individually these modes are not squeezed and resemble a thermal state
\cite{barnett1985,yurke1987,barnett1988}. 
Due to the correlations and symmetry between the two modes, the average photon number in each mode is the same. Also, the covariance between the two modes describe the inter-mode correlations. TMSV is produced experimentally via non degenerate parametric downconversion or four-wave mixing
\cite{GK,Slusher1985}. 
Recently TMSV light has proven to be extremely useful in quantum metrology
\cite{dowling2010, dowling2017}
and quantum information processing 
\cite{cochrane2000}. 

The paper is organized as follows. In section \RN{1}, we propose the scheme to resolve photon number at room temperature without using photon-number-resolving detectors, and calculate the output signal. We use a two-mode squeezing device, such as an optical parametric amplifier (OPA), or a four-wave mixer (FWM), in a spatially non degenerate configuration. In section \RN{3}, we analyze our scheme in the presence of losses, and calculate the signal-to-noise ratio.
%A $N$-photon Fock state (to be detected at the output) and coherent light are used as inputs to the FWM. At the output, coincidence counts are measured. The average coincidence counts $\langle \hat{C} \rangle$ and the noise in the coincidence counts $\Delta \hat{C}$ are used to detect the input photon number. The coincidence count calculated in this work is actually the intensity-intensity correlation function. Both coincidence count and intensity-intensity correlation are defined by the same operator $\langle\hat{a}^{\dagger}\hat{a}\hat{b}^{\dagger}\hat{b}\rangle$.
%One of the challenges of using devices based on parametric downconversion is matching the phases between the input light sources and the pump. However, in our technique, the phase cancels out due to the presence of Fock states, making it easy to implement in the lab (this would also be the case if the signal were a weak thermal state).
\section{\label{sec:level2}PHOTON-NUMBER RESOLVING SCHEME}
\begin{figure}
 \centering
 \includegraphics[scale=0.45]{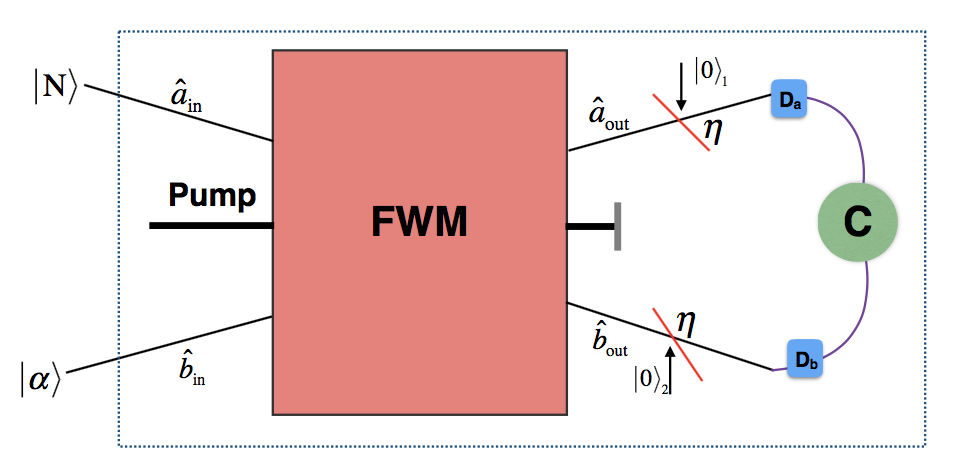}
\caption{The schematic diagram of a room-temperature number-resolving photon detector. The two-mode inputs to the four-wave mixer (FWM) are $N$-photon Fock states, and a coherent state of light $|\alpha\rangle$, $\hat{a}_{\mathrm{in}} (\hat{a}_{\mathrm{out}})$ and $\hat{b}_{\mathrm{in}} (\hat{b}_{\mathrm{out}})$ represent the mode operators of input (output) light beams. The average intensity-intensity correlations and the noise in the intensity-intensity correlations are detected at the output. The losses due to imperfect squeezing and the inefficiency of the photon detectors, are modeled by adding fictitious beam splitters each of overall transmissivity $\eta$, where the vacuum modes are denoted by $|0\rangle_{1}$ and $|0\rangle_{2}$.}
\label{fwm}
\end{figure}
The setup used for the proposed scheme is shown in Fig.~\ref{fwm}. An unknown $N$-photon state is incident on one port of the FWM and a coherent-light state with average photon number $\bar{n}_{\alpha}$ is incident on the second port. The average intensity-intensity correlations $\langle \hat{C} \rangle$ and the noise in the intensity-intensity correlations $\Delta \hat{C}$ are measured at the output to detect the input photon number. The operators $\hat{a}$ and $\hat{b}$ after interacting with the two-mode squeezer become
\begin{eqnarray}
\hat{a} \rightarrow \hat{a}\mu-\hat{b}^{\dagger}\nu \nonumber\\
\hat{b} \rightarrow \hat{b}\mu-\hat{a}^{\dagger}\nu,
\end{eqnarray} 
where $\mu=\cosh(r)$, and $\nu=\sinh(r)$.
Intensities $\hat{N}_{a}$ and $\hat{N}_{b}$ and the intensity difference $\langle\hat{M^{ab}_{-}}\rangle$ at the two output modes are 
%\begin{eqnarray}
\begin{alignat}{3}
\langle\hat{N}_{a}\rangle  =  \bar{n}_{s}(\bar{n}_{\alpha}+N) +N+\bar{n}_{s}, \nonumber \\
\langle\hat{N}_{b}\rangle  = \bar{n}_{s}(\bar{n}_{\alpha}+N) +\bar{n}_{\alpha}+\bar{n}_{s},\nonumber \\
\langle\hat{M}^{ab}_{-}\rangle = N-\bar{n}_{\alpha},
\end{alignat}
%\end{eqnarray}
where $\bar{n}_{s}$ is the average number of photons in a single-mode squeezed vacuum and is fixed at the value of two in this calculation, corresponding to 10 dB of squeezing 
\cite{tobias2013,lvovsky2015}.
\begin{figure}
\centering
  \includegraphics[scale=0.57]{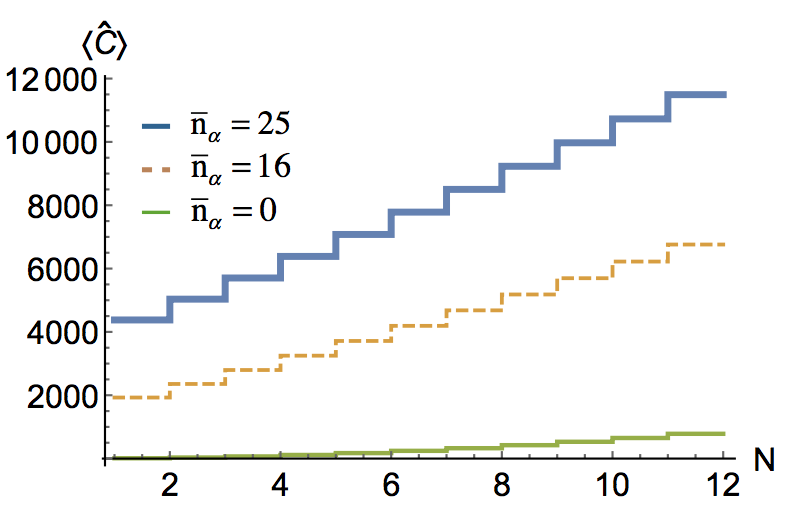}\vfil
   \includegraphics[scale=0.61]{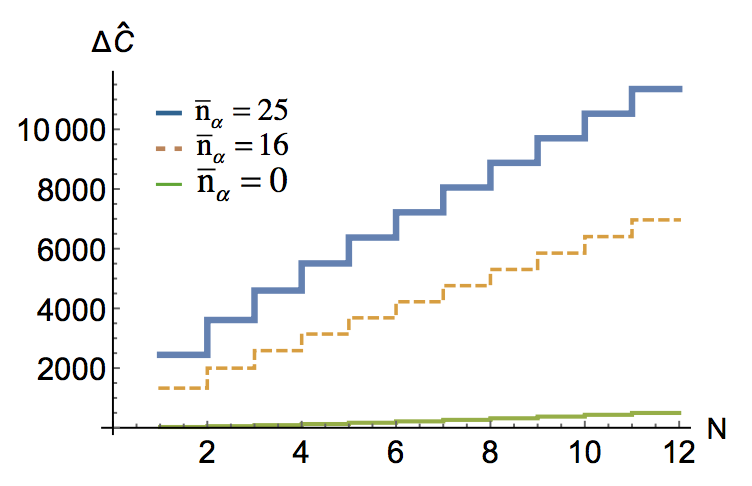}
\caption{(a) and (b) show the average intensity-intensity correlations $\langle\hat{C} \rangle$ and the noise $\Delta\hat{C}$ in the intensity-intensity correlations as a function of input photon number $N$ incident on one port of a two-mode squeezing device with $\bar{n}_{s}=2$ respectively: Both $\langle\hat{C} \rangle$ and $\Delta\hat{C}$ increase in steps as the input photon number changes in increments of one. When a single photon is incident, there is huge jump in $\langle\hat{C} \rangle$ and $\Delta\hat{C}$. $\langle\hat{C} \rangle$ and $\Delta\hat{C}$ for vacuum as input in the second mode shows smaller step sizes than those with coherent-light inputs. Hence the coherent-light state provides a $\textit{boost}$ to the $\langle\hat{C} \rangle$ and $\Delta\hat{C}$ signals. Also this shows that even in the presence of coherent state amplitude fluctuation, we still see the steps in the signal and the noise. Therefore, for a slowly fluctuating coherent state, we expect to observe slowly fluctuating signal while still maintaining the steps, representing the input photon number.
} %In this work we call $\Delta C$ as the signal due to the large magnitude of the jump on the detection of a single photon. 
\label{coinc}
\end{figure}
The above equations show that correlations and symmetry between the two modes has been disturbed because of different input modes. In particular $\langle\hat{M}^{ab}_{-}\rangle$ is identically zero for pure TMSV. We exploit this change in the correlations between the two beams to resolve the number of photons in the input by detecting the average intensity-intensity correlations at the output. The average intensity-intensity correlations signal is calculated from
\begin{equation}
%\label{ }
\langle\hat{C} \rangle = \langle N |_{a}\langle \alpha |_{b} \hat{N}_{a}\hat{N}_{b} | \alpha\rangle_{b}|N\rangle_{a}, 
\end{equation}
and is given by 
%eq.~\ref{coincidence}.
\begin{align}
\label{coincidence}
%\begin{split}
\langle\hat{C}\rangle&= \left(\alpha ^2+1\right) (N+1) \sinh ^4(r)\nonumber\\
&+\alpha ^2 N (1+\sinh^2(r))^2 \nonumber\\
&+\left(\alpha ^4+\alpha^2 (2 N+3)+(N+1)^2\right) \sinh ^2(r)\nonumber\\
& (1+\sinh^2(r)).
%   \end{split}
\end{align}
\begin{figure}
\begin{center}
\includegraphics[scale=0.6]{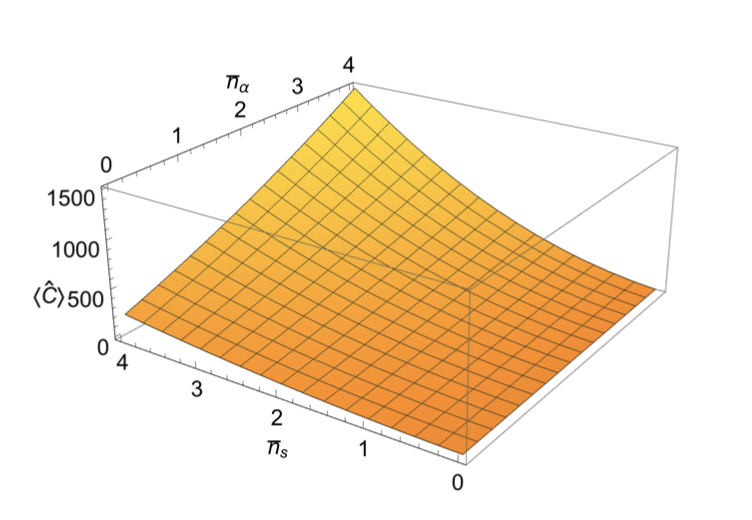}
\caption{The average intensity-intensity correlations signal as a function of $\bar{n}_{\alpha}$, and $\bar{n}_{s}$. The signal attains the maximum value at $\bar{n}_{\alpha}=\bar{n}_{s}$.}
\label{optimum}
\end{center}
\end{figure}
\begin{figure}
\begin{center}
\includegraphics[scale=0.5]{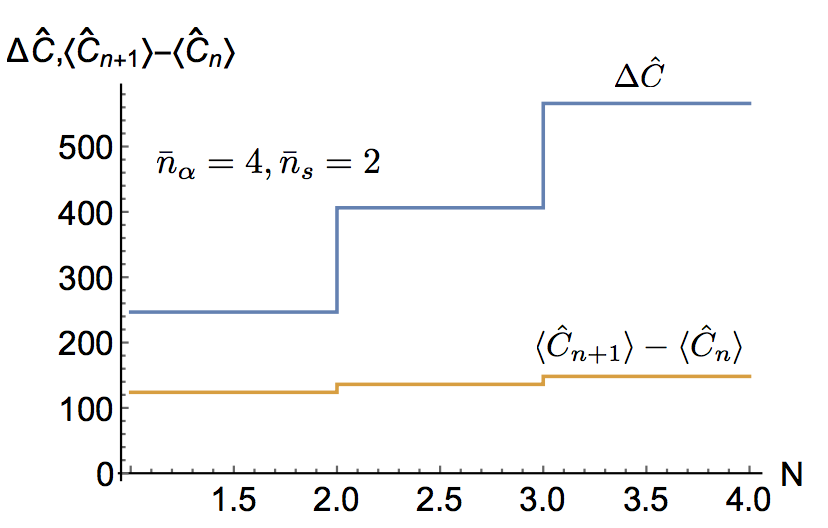}
\caption{Plot comparing the step-size of the average intensity-intensity correlations signal, with the noise.}
\label{stepsize}
\end{center}
\end{figure}

\begin{figure}
\begin{center}
\includegraphics[scale=0.45]{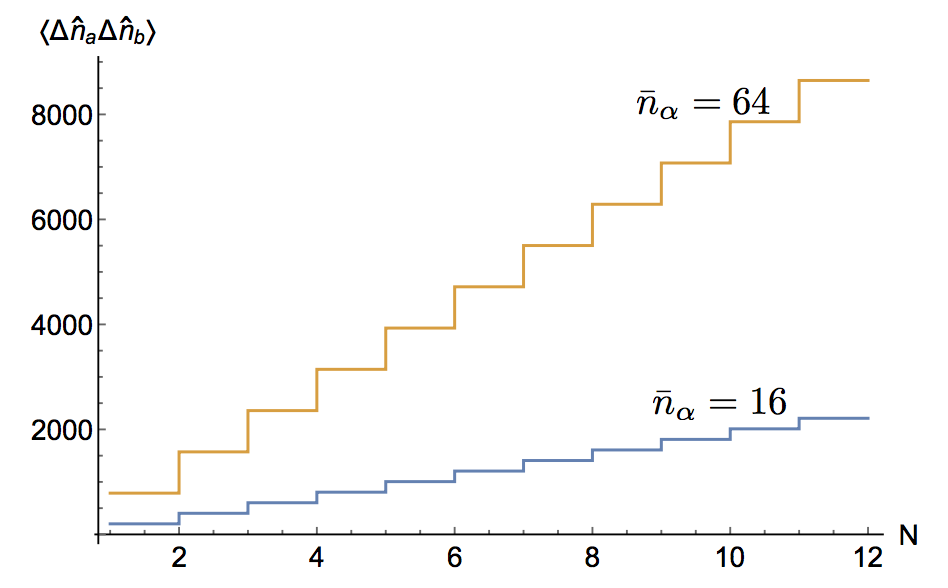}
\caption{Correlation in photon number fluctuations as function of input photon number.}
\label{covariance}
\end{center}
\end{figure}

The average intensity-intensity correlations and the standard deviation (noise) of the average intensity-intensity correlations as a function of the input photon state are plotted in Fig.~\ref{coinc}, (see appendix for the expression of $\Delta \hat{C}$). 
From the figure we can see that there is a huge jump in both $\langle \hat{C} \rangle$ and $\Delta \hat{C}$ even when a single photon is incident on the FWM. What is interesting is the amplification of the noise in the  intensity-intensity correlations when a single photon is detected. Hence, a large change in $\Delta \hat{C}$ is an indicator of the presence of photon. 
In Fig.~\ref{coinc} we compare the amplitude of the signal for vacuum and coherent-light input respectively. The steps for the case of nonzero coherent-light input are greatly amplified compared to the vacuum, and hence this provides a boost to the intensity-intensity correlations signal. Thus the purpose of having coherent light as input to the second mode is to amplify the output signal while still displaying the steps as the photon number changes. Our scheme does not require very strong coherent light source, therefore the possibility of the coherent-light producing its own twin beam state is ruled out. In order to have a well calibrated non-linear gain, a feedback system to control the output measured coherent-state amplitude can be used. This will be equivalent to controlling the gain, while showing the jumps in the $\langle \hat{C} \rangle$ or $\Delta \hat{C}$ signals.
%However, we would like to remark that in the regime of large $\bar{n}_{\alpha}$, $\langle\hat{C}\rangle > \Delta\hat{C}$ (see supplementary material). 
Both $\langle \hat{C} \rangle$ and $\Delta \hat{C}$ display steps as the number of input photon is increased in steps of one. Therefore, it is possible to know the input photon number by counting the height of steps in $\langle \hat{C} \rangle$ or $\Delta \hat{C}$. 
In Fig.~\ref{optimum} we show that the $\langle\hat{C}\rangle$ signal is maximum when both $\bar{n}_{\alpha}$, and $\bar{n}_{s}$ are equal. Also, both the $\langle\hat{C}\rangle$, and $\Delta\hat{C}$ are comparable in magnitude for any choice of $\bar{n}_{\alpha}$, and $\bar{n}_{s}$. Therefore, the step size of $\langle\hat{C}\rangle$ signal can never exceed the noise, $\Delta\hat{C}$. Hence, the current set-up is not suitable for single shot experiment. In Fig.~\ref{stepsize} we compare the noise, and the step-size. We can also use the covariance or the correlation in photon number fluctuations as a function of input photon number, shown in Fig.~\ref{covariance} as the signal.

\begin{figure}
\begin{center}
\includegraphics[scale=0.6]{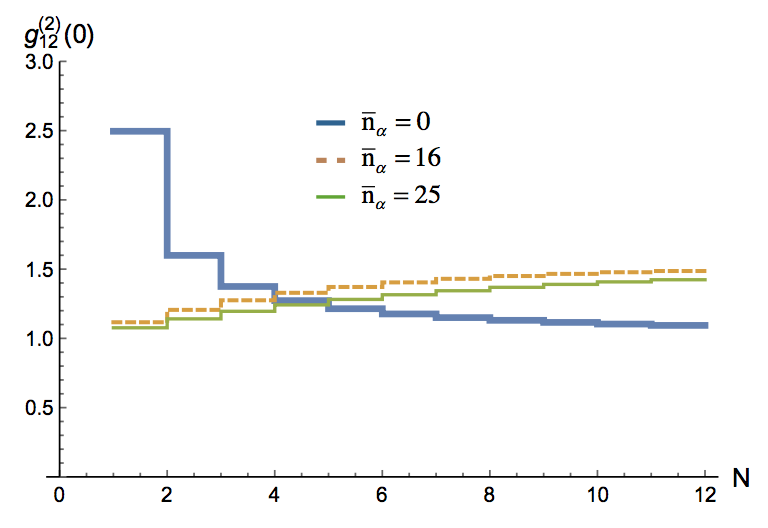}\vfil
 \includegraphics[scale=0.6]{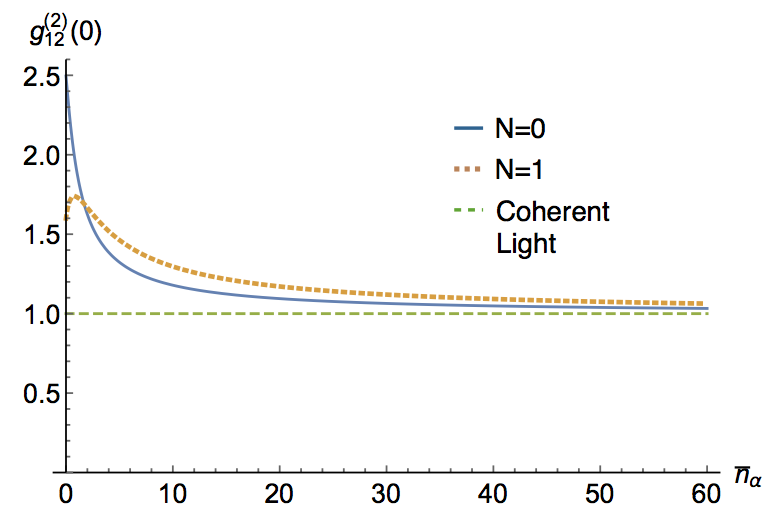}
\caption{(a) Two-mode second-order intensity correlation function $g_{12}^{(2)}(0)$ as a function of number of input photons $N$ for different $\bar{n}_{\alpha}$, and $\bar{n}_{s}=2$. When $\bar{n}_{\alpha} = 0$, then $g_{12}^{(2)}(0)$ has maximum correlation for $N$=0. As $N$ increases, $g_{12}^{(2)}(0)$ decreases. When $\bar{n}_{\alpha} \neq 0$, the correlations increase with $N$, but still less than that of TMSV. (b) $g_{12}^{(2)}(0)$ as a function of coherent-light amplitude. As the strength of the coherent light is increased  the curves for $N=0$, and $N=1$ approach the single-mode second-order intensity correlation function $g_{1}^{(2)}(0)$ for a coherent state asymptotically.}
\label{g12}
\end{center}
\end{figure}

We also calculate the two-mode second-order intensity correlation function $g_{12}^{(2)}(0)$ which is defined as $\langle\hat{C}\rangle/(\langle\hat{N}_{a}\rangle\langle\hat{N}_{b}\rangle)$
\cite{Milburn},
and is calculated at zero time delay. 
This is another way of describing the intermode correlations as well as photon bunching. We know that if $g_{12}^{(2)}(0) > 1$, then the light has bunching or represents super-Poisson state. For a two-mode squeezed vacuum light, $g_{12}^{(2)}(0)$ $=$ $2+1/\bar{n}_{s}$, where $\bar{n}_{s}$ is the average photon number in a single-mode squeezed vacuum state. The  $g_{12}^{(2)}(0)$ for $|N\rangle_{a} \otimes |\alpha\rangle_{b}$ input is 
 \begin{widetext}
\begin{equation}
\label{gfn}
g_{12}^{(2)}(0) = \frac{N\bar{n}_{\alpha}(\bar{n}_{s}^{2}+(1+\bar{n}_{s})^{2})+((N+1)^{2}+(2N+3)\bar{n}_{\alpha}+\bar{n}_{\alpha}^{2})\bar{n}_{s}(1+\bar{n}_{s})+(1+N+\bar{n}_{\alpha})\bar{n}_{s}^{2}}{(\bar{n}_{\alpha}(1+\bar{n}_{s})+(N+1)\bar{n}_{s})(N(1+\bar{n}_{s})+(1+\bar{n}_{\alpha})\bar{n}_{s})}.
\end{equation}
\end{widetext}
In Fig.~\ref{g12}.a we plot $g_{12}^{(2)}(0)$ as a function of the coherent-state mean photon number $\bar{n}_{\alpha}$. As the strength of input coherent light increases, the correlations between the two modes decreases and $g_{12}^{(2)}(0)$ approaches the single-mode second-order intensity correlation function $g^{(2)}(0)$) of a coherent-light state, asymptotically.  Also, we see that the presence of a single photon in the input mode is sufficient to reduce the correlations between the two beams.
%%%%%%%%%Efficiency Calculation%%%%%%%%
\section{\label{sec:level3}Effect of Losses}
\begin{figure}
\begin{center}
\includegraphics[scale=0.4]{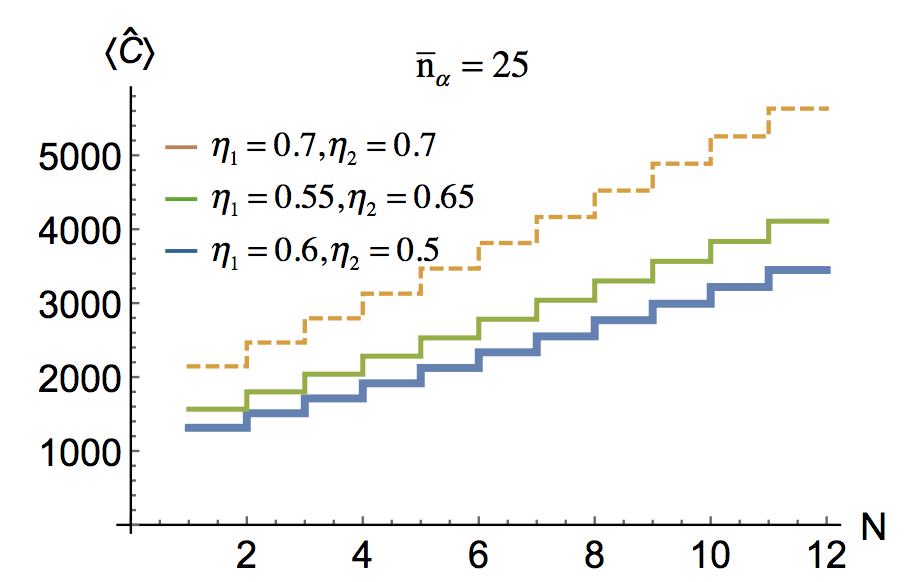}\hfil
\includegraphics[scale=0.4]{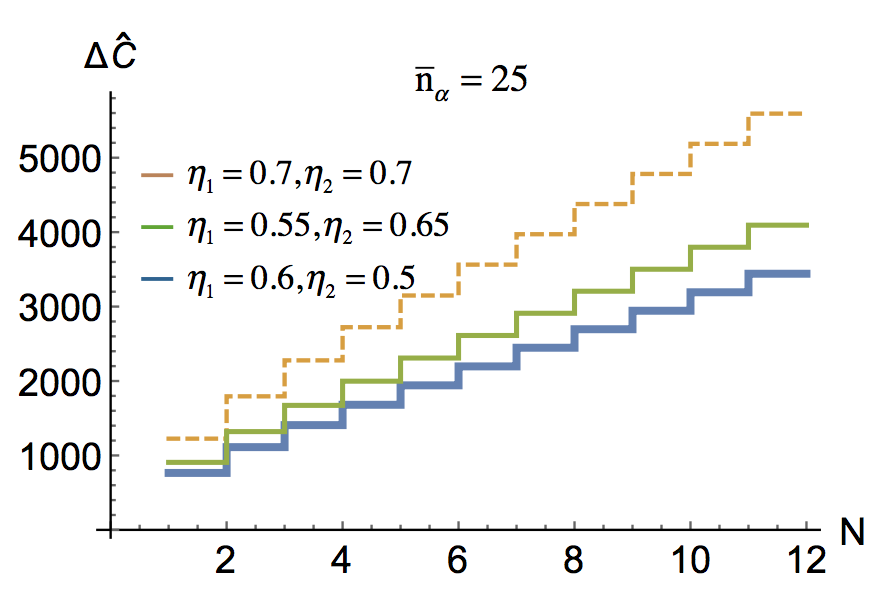}\hfil
\includegraphics[scale=0.4]{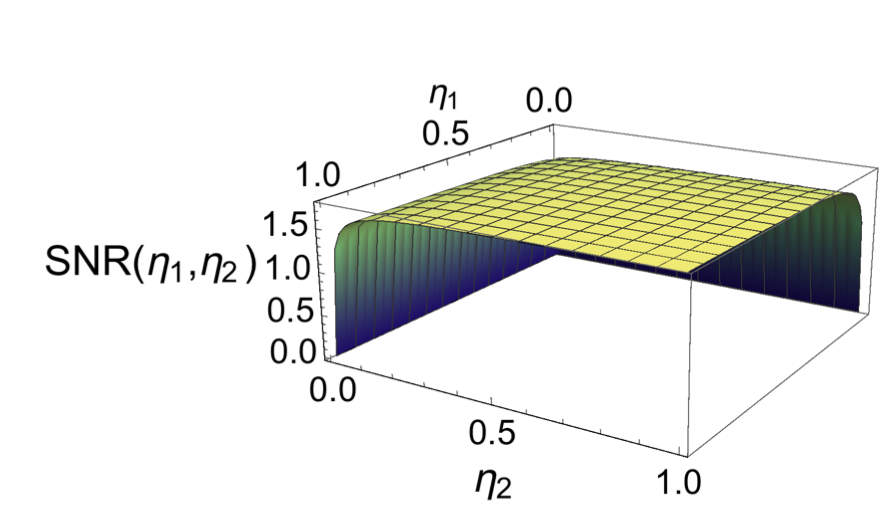}
\caption{(a) and (b) $\langle\hat{C}\rangle$ and $\Delta\hat{C}$ as a function of $\eta_{1}$, and $\eta_{2}$, plotted against the number of input photons for $\bar{n}_{\alpha} = 25$. (c) SNR plotted against $\eta_{1}$, and $\eta_{2}$.}
\label{T1T2}
\end{center}
\end{figure}

%\subsection{\label{sec:level1}Model: \texttt{reprint}}
Next, we address the issue of imperfect squeezing and inefficient detection of photons. Generally, the devices used to produce two-mode squeezed light do not perform perfect squeezing and the TMSV is a mixed state. Also, the photon detectors used to detect the photons also have a limited efficiency leading to losses. We model these losses by introducing fictitious beam splitters of transmissivity $\eta=\epsilon T$, where $\epsilon$ represents imperfect squeezing and $T$ represents the efficiency of the photon detectors. Therefore the total loss is $1-\eta$. 
\begin{figure}
\begin{center}
\includegraphics[scale=0.6]{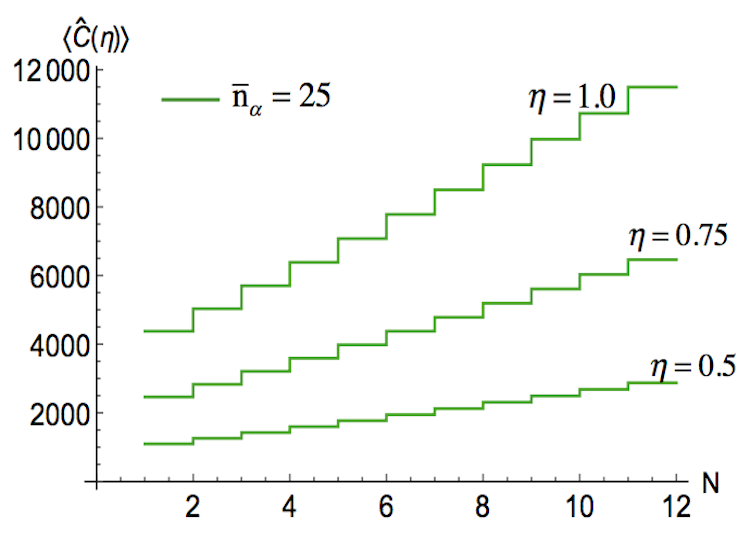}\vfil
 \includegraphics[scale=0.6]{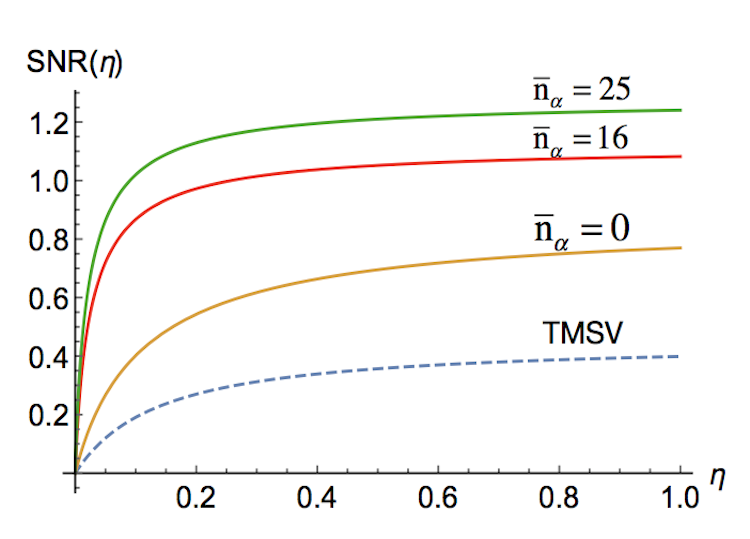}
\caption{(a) Average intensity-intensity correlations signal as a function of input photon number, for different efficiencies represented by $\eta$ and fixed $\bar{n}_{\alpha}=25$. The imperfect two-mode squeezing and correlator can be modeled by adding fictitious beam splitters of transmissivity defined as $\eta=\epsilon T$. Where $\epsilon$ represents imperfect squeezing and $T$ represents the efficiency for the photon detector. (b) Signal-to-noise ratio (SNR) as a function of $\eta$.}
\label{effT}
\end{center}
\end{figure}
Fig.~\ref{effT}.a shows $\langle\hat{C}\rangle$ in the presence of losses as a function of the input photon number $N$. We can see that as the efficiency increases the amplitude of the signal $\langle\hat{C}\rangle$ increases. Also it is possible to attain the same amplitude of the intensity-intensity correlation signal even when the efficiency is low ($\eta \sim 0.5$), by using a stronger coherent-light source to compensate. Hence, the use of coherent light acts as a boost that overcomes the effect of inefficiency in the squeezing and photon detection. 
Also, unbalanced detector inefficiencies and losses ($\eta_{1} \neq \eta_{2}$) frequently give rise to adverse effects in experimental quantum optics schemes.  However, in our scheme, having detectors of different efficiencies does not degrade the signal, nor the performance, of the PNRD. In Fig.~\ref{T1T2} we plot the signal-to-noise ratio as a function of $\eta_{1}$, and $\eta_{2}$ and the average intensity-intensity correlations  when the two detector efficiencies are different. 
Additionally, phase-sensitive detection and amplification schemes are difficult to implement experimentally, as care must be taken to control the (typically) optical phases of the involved beams. Our scheme avoids such difficulties, as the relative phases of the involved modes is not an issue due to the orthogonality of the Fock states. This is true for the thermal state as well, so we can use our scheme to detect weak thermal light. However, it is worth noting that this does not overrule the mode-matching with respect to to the wave-vectors between the different input modes to complete the non-linear process.
The signal-to-noise ratio (SNR) is a measure of the system performance. It is defined as,  
\begin{eqnarray}
%SNR  = \Delta \hat{C}/\langle\hat{C} \rangle.   
\mathrm{SNR}  =\langle\hat{C} \rangle/ \Delta \hat{C}.   
\end{eqnarray}
In Fig.~\ref{effT}.b we plot the signal-to-noise ratio as a function of the transmissivity (see appendix for the expression of SNR). The SNR decreases as the transmissivity decreases, however this can be compensated for by increasing the strength of the coherent-light state. 

\begin{figure}
\begin{center}
\includegraphics[scale=0.5]{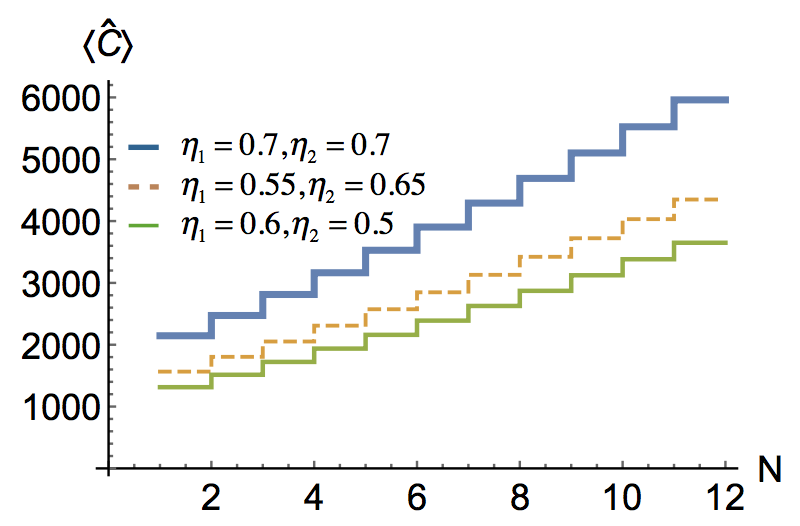}\hfil
\includegraphics[scale=0.4]{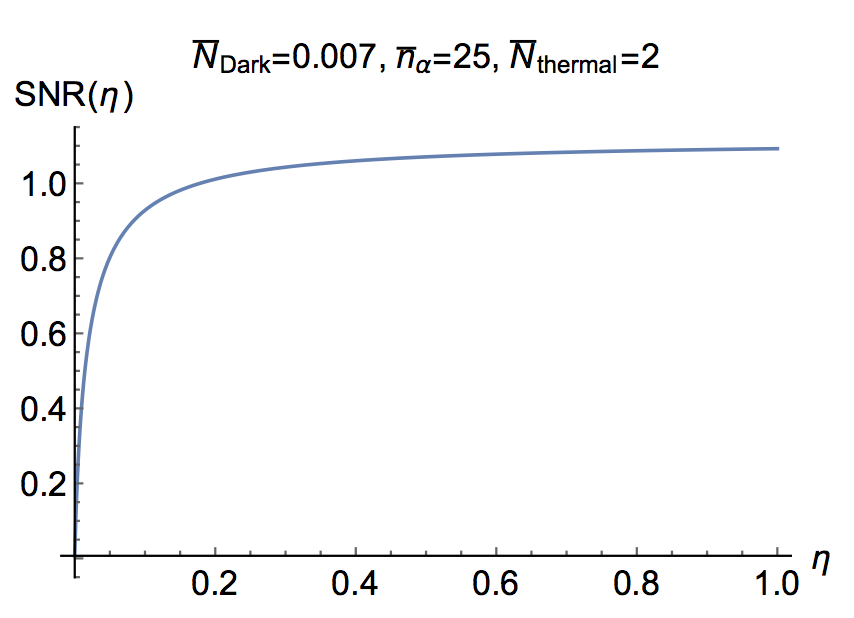}
\caption{(a)  Average intensity-intensity correlations signal as a function of input photon number, for different efficiencies represented by $\eta$ and fixed $\bar{n}_{\alpha}=25$. Again, the \textit{N} used in this plot is the $\overline{N}_{\mathrm{thermal}}$. (b) Signal-to-noise ratio in the presence of dark counts against detector efficiency.}
\label{snrDark}
\end{center}
\end{figure}

We also address the effect of stray thermal photons on our detection scheme. The thermal photons at room temperature are completely uncorrelated between detectors, and the average number of photons at optical frequencies is very small ($\sim 10^{-40}$)
\cite{GK}. 
Again the average number of stray thermal photons at room temperature is of the order of $10^{-3}$, which does not effect the detector efficiency.
\begin{figure}
\begin{center}
\includegraphics[scale=0.5]{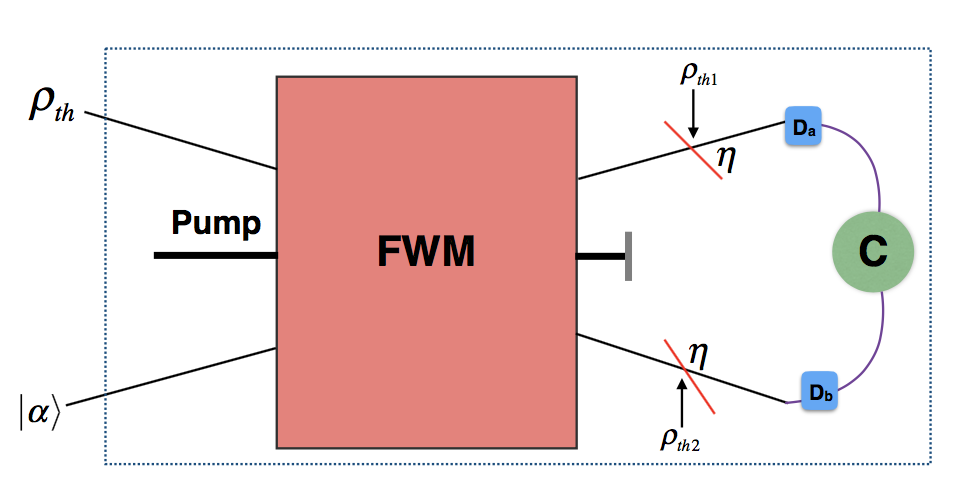}
\caption{The effect of dark counts on the room-temperature number-resolving photon detector. The input number state is approximated with thermal state. The losses due to imperfect squeezing and the dark counts at the output, are modeled by adding fictitious beam splitters each of overall transmissivity $\eta$, where the thermal modes are denoted by $\rho_{\mathrm{th1}}$ and $\rho_{\mathrm{th2}}$.}
\label{fwmthermal}
\end{center}
\end{figure}
\begin{figure}
\begin{center}
\includegraphics[scale=0.5]{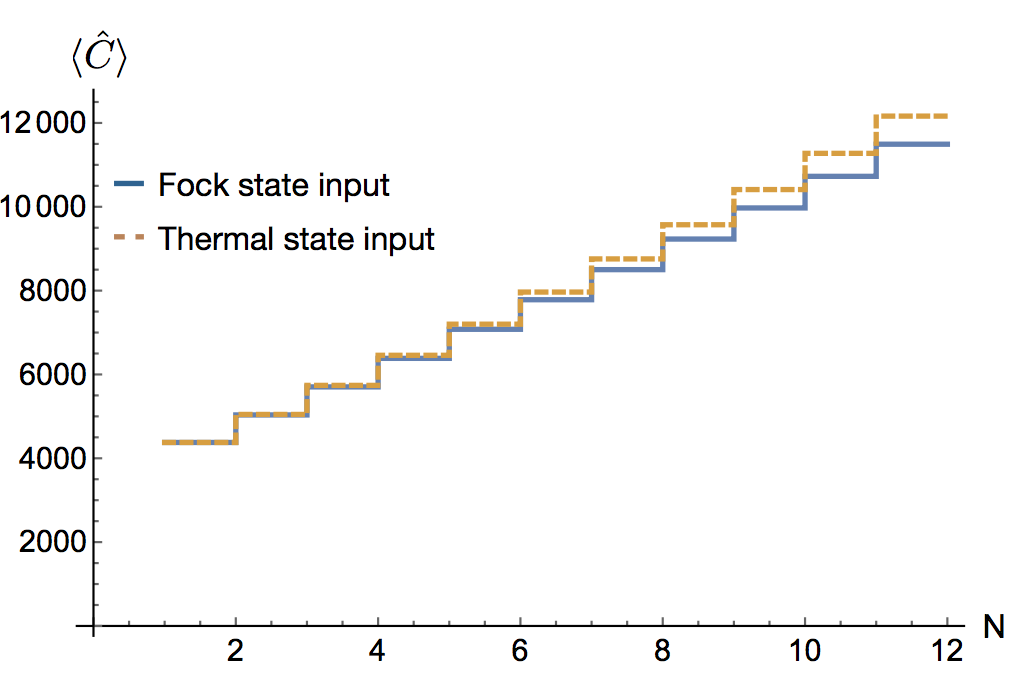}
\caption{Plot comparing the average intensity-intensity correlations as a function of number of input photons \textit{N} for a Fock state input, and a thermal state input. For a thermal input state, \textit{N} is actually $\overline{N}_{\mathrm{thermal}}$, which we have chosen to be an integer increasing in increments of one. This helps in making an easier comparison between the two input states. The average number of photons in the coherent state is $\bar{n}_{\alpha}=25$. From the plot we can see that $\langle\hat{C}\rangle$ does not vary much for the two different input states. Hence, we can conclude that the thermal state is a good approximation for input Fock state in the calculation for dark counts.}
\label{Fock_thermal_cmp}
\end{center}
\end{figure}
We mix the stray thermal photons with the output at the two beam splitters as shown in Fig.~\ref{fwmthermal}. In order to make the dark count calculation easier, we approximate the Fock states with a thermal state enabling us to use Wigner functions
\cite{bryanthesis}.
We compare the average intensity-intensity correlations between the input Fock state and the thermal state input in Fig.~\ref{Fock_thermal_cmp}. We find that the the two signals do not differ much, hence the the thermal state is a good approximation for the Fock state as input, and we expect the effect of dark counts on an actual number state $|N\rangle$ to be similar.  
In Fig.~\ref{snrDark} we show the effect of stray thermal photons on the intensity-intensity correlations signal and the signal-to-noise ratio. The average number of thermal photons $\overline{N}_{\mathrm{Dark}}$ at the room temperature i.e. 300K, have been calculated at the wavelength of 9.7$\mathrm{\mu}$m.
%\section{\label{sec:level4}Applications} 
%Describe photon statistics for source characterization.
%QKD
%Thermal imaging-weal thermal light detection
%%%%%%Conclusion%%%%%%%%
\section{\label{sec:level5}Conclusion}
In summary we propose a room-temperature photon-number-resolving detector using a two-mode squeezer. The $N$-photon number state is fed into a two-mode squeezing device, along with a coherent-light input which amplifies the output signal. The output intensity-intensity correlations signal reports jumps with the changing photon number. Even in the presence of losses, the output signal is strong due to the amplification provided by the coherent light. Hence, we have a high efficiency photon-number-resolving detector. Since the scheme is robust against low detector efficiency, the intensity-intensity correlation measurement can be carried out at room temperature for optical photons. 

Additionally since the photon-number states to be counted are boosted (amplified) in the squeezer, dark counts will have negligible effect, particularly at room temperature. Also, this particular setup is robust against any phase fluctuations due to the presence of Fock states which are insensitive to phase. Hence, phase matching is not required, making our technique easier to implement in the lab. 
Also, the synchronization of the different light pulses will depend mainly on the coherent state. Most experiments use a continuous-wave coherent light which will give a steady background signal, and is easy to synchronize due to a narrower line width. Moreover, if the temporal profile of the input Fock state is known, it is easy to produce coherent light with the same temporal profile. Also our scheme is robust against coherent state amplitude fluctuation, as we still see the steps in the signal and the noise. Therefore, for a slowly fluctuating coherent state, we expect to observe slowly fluctuating signal but still maintaining the steps, representing the input photon number.
Since, both $\langle\hat{C} \rangle$, and $\Delta \hat{C}$ are comparable in magnitude, the step-size never exceeds the noise, $\Delta \hat{C}$. Therefore, the current set-up is not suitable for a single-shot experiment. Our results can be applied to a wide range of squeezers and each would need to be addressed separately in any experiment. Similarly, the time required for ensemble measurements would depend on the different experiments.  

Our scheme is not a general photon-number-resolving detector because it does not implement the POVM $|N\rangle \langle N|$ in the $|N\rangle$ basis. Therefore for thermal light, squeezed light, and coherent light, it will give a distribution around the mean. However, for many applications in quantum technology such as 
%Boson sampling
%\cite{rhode2015b},
quantum key distribution
\cite{sumeet2017},
the photon state is  known to be in a Fock state, which is unknown. For such applications our scheme will be ideal. Nevertheless, because of the coherent light boosting, this device should be useful for detecting weak thermal light, squeezed light, and coherent light states that has application for example in quantum LIDAR
\cite{dowling2013}. In future work, we plan to explore our setup for multi-frequency mode.
%%%%%%Aknowledgement%%%%%%
\begin{acknowledgments}
The authors would like to acknowledge the Air Force Office of Scientific Research, the Army Research Office, the Defense Advanced Research Projects Agency, the National Science Foundation, and the Northrop Grumman Corporation. Also, the authors would like to thank Hagai Eisenberg, Lior Cohen from the Hebrew University of Jerusalem, and Vincenzo Tamma from the University of Portsmouth for enriching discussions.
\end{acknowledgments}
%%%%%%Bibliography%%%%%

\begin{widetext}
\section*{Appendix A}
The average intensity-intensity correlation for imperfect detections with efficiency $\eta$ is
\begin{align}
\langle\hat{C} (\eta) \rangle&= \eta^2 (N \alpha^2 (1+\sinh^2(r))^2 + ((1 + N)^2 + (3 + 2 N) \alpha^2 + \alpha^4)(1+\sinh^2(r)) \sinh^2(r) \nonumber \\
&+ (1 + N) (1 + \alpha^2) \sinh^4(r)).
\end{align}
The expressions for the variance in intensity-intensity correlation signal $\Delta\hat{C}^{2} = \langle\hat{C}^{2}\rangle- \langle\hat{C}\rangle^{2}$ and signal- to-noise ratio are given by the following equations,
\begin{align}
\Delta\hat{C}^{2}&= N^2 \alpha^2 (1 + \sinh(r)^2)^4 + (1 +  \sinh(r)^2)^3  \sinh(r)^2 + 3 N (1 +\sinh(r)^2)^3  \sinh(r)^2 \nonumber \\
&+ 3 N^2 (1 +  \sinh(r)^2)^3  \sinh(r)^2+ N^3 (1 +  \sinh(r)^2)^3  \sinh(r)^2  \nonumber \\
&+ 7 \alpha^2 (1 +  \sinh(r)^2)^3  \sinh(r)^2 +19 N \alpha^2 (1 + \sinh(r)^2)^3 \sinh(r)^2 \nonumber \\
&+ 11 N^2 \alpha^2 (1 + \sinh(r)^2)^3 \sinh(r)^2 + 2 N^3 \alpha^2 (1 + \sinh(r)^2)^3 \sinh(r)^2+ 6 \alpha^4 (1 + \sinh(r)^2)^3 \sinh(r)^2\nonumber \\
&+ 13 N \alpha^4 (1 + \sinh(r)^2)^3 \sinh(r)^2 +4 N^2 \alpha^4 (1 + \sinh(r)^2)^3 \sinh(r)^2 + \alpha^6 (1 + \sinh(r)^2)^3 \sinh(r)^2 \nonumber \\
&+ 2 N \alpha^6 (1 + \sinh(r)^2)^3 \sinh(r)^2 + 10 (1 + \sinh(r)^2)^2 \sinh(r)^4 + 20 N (1 + \sinh(r)^2)^2 \sinh(r)^4 \nonumber \\
&+12 N^2 (1 + \sinh(r)^2)^2 \sinh(r)^4+ 2 N^3 (1 + \sinh(r)^2)^2 \sinh(r)^4 + 43 \alpha^2 (1 + \sinh(r)^2)^2 \sinh(r)^4\nonumber \\
&+ 68 N \alpha^2 (1 + \sinh(r)^2)^2 \sinh(r)^4 + 32 N^2 \alpha^2 (1 + \sinh(r)^2)^2 \sinh(r)^4 \nonumber \\
&+4 N^3 \alpha^2 (1 + \sinh(r)^2)^2 \sinh(r)^4 + 32 \alpha^4 (1 + \sinh(r)^2)^2 \sinh(r)^4 \nonumber \\
&+ 36 N \alpha^4 (1 + \sinh(r)^2)^2 \sinh(r)^4 + 10 N^2 \alpha^4 (1 + \sinh(r)^2)^2 \sinh(r)^4 \nonumber \\
&+ 6 \alpha^6 (1 + \sinh(r)^2)^2 \sinh(r)^4 + 4 N \alpha^6 (1 + \sinh(r)^2)^2 \sinh(r)^4 + 9 (1+ \sinh(r)^2) \sinh(r)^6 \nonumber \\
&+ 15 N (1 + \sinh(r)^2) \sinh(r)^6 + 7 N^2 (1 + \sinh(r)^2) \sinh(r)^6 + N^3 (1 + \sinh(r)^2) \sinh(r)^6  \nonumber \\
&+ 29 \alpha^2 (1 + \sinh(r)^2) \sinh(r)^6+ 47 N \alpha^2 (1 + \sinh(r)^2) \sinh(r)^6 + 19 N^2 \alpha^2 (1 + \sinh(r)^2) \sinh(r)^6 \nonumber \\
&+ 2 N^3 \alpha^2 (1 + \sinh(r)^2) \sinh(r)^6+ 14 \alpha^4 (1 + \sinh(r)^2) \sinh(r)^6 \nonumber \\
&+ 21 N \alpha^4 (1 + \sinh(r)^2) \sinh(r)^6 + 4 N^2 \alpha^4 (1 + \sinh(r)^2) \sinh(r)^6 \nonumber \\
&+\alpha^6 (1 + \sinh(r)^2) \sinh(r)^6 + 2 N \alpha^6 (1 + \sinh(r)^2) \sinh(r)^6\nonumber \\
&+\alpha^2 \sinh(r)^8 + 2 N \alpha^2 \sinh(r)^8 + N^2 \alpha^2 \sinh(r)^8, 
\label{eq:delC}
\end{align}
\begin{align}
\mathrm{SNR}&=(\eta^2 (N \alpha^2 (1+\sinh(r)^2)^2 + ((1 + N)^2 + (3 + 2 N) \alpha^2 + \alpha^4) (1+\sinh(r)^2) \sinh(r)^2 \nonumber \\
&+ (1 + N) (1 + \alpha^2) \sinh(r)^4))/  \nonumber \\
&((1 -\eta)^2\eta^2 (N \alpha^2 (1+\sinh(r)^2)^2 + ((1 + N)^2 + (3 + 2 N) \alpha^2 + \alpha^4) (1+\sinh(r)^2) \sinh(r)^2 \nonumber \\
&+(1 + N) (1 + \alpha^2) \sinh(r)^4)-\eta^4 (N \alpha^2 (1+\sinh(r)^2)^2 \nonumber\\
&+ ((1 + N)^2 + (3 + 2 N) \alpha^2 + \alpha^4) (1+\sinh(r)^2) \sinh(r)^2 \nonumber \\
&+ (1 + N) (1 + \alpha^2) \sinh(r)^4)^2 +(1 - \eta) \eta^3 (N (\alpha^2 + \alpha^4) (1+\sinh(r)^2)^3 + (2 \alpha^2 + N^2 \alpha^2 \nonumber \\
&+ 2 N (1 + N) \alpha^2 + 4 \alpha^4 + \alpha^6 + N (1 + N) (1 + \alpha^2) + N (\alpha^2 + 2 \alpha^4) \nonumber \\
&+ (1 + N) (1 + 5 \alpha^2 + 2 \alpha^4)) (1+\sinh(r)^2)^2 \sinh(r)^2 + (N (1 + N)^2 + (N^2 + N (1 + N)) \alpha^2 \nonumber \\
&+ (1 + N) (3 + 2 N) (1 + \alpha^2)+ N (2 \alpha^2 + \alpha^4)  \nonumber \\
&+ (1 + N) (1 + 7 \alpha^2 + 3 \alpha^4)) (1+\sinh(r)^2) \sinh(r)^4 + (1 + N)^2 (1 + \alpha^2) \sinh(r)^6) \nonumber \\
&+ (1 - \eta) \eta^3 (N^2 \alpha^2 (1+\sinh(r)^2)^3 + (N^2 (1 + N) + (-1 + N) N \alpha^2 + N^2 \alpha^2 +N (1 + N) (1 + \alpha^2) \nonumber \\
&+ (1 + N)^2 (1 + \alpha^2) + N (\alpha^2 + \alpha^4) + 2 N (2 \alpha^2 + \alpha^4) + (1 + N) (2 \alpha^2 + \alpha^4)) (1+\sinh(r)^2)^2 \sinh(r)^2 \nonumber \\
& + (4 \alpha^2 + N (1 + N) \alpha^2 + 5 \alpha^4 + \alpha^6 + 2 N (1 + N) (1 + \alpha^2) + (1 + N)^2 (1 + \alpha^2) \nonumber \\
&+ N (\alpha^2 + \alpha^4) + N (2 \alpha^2 + \alpha^4) + (1 + N) (1 + 3 \alpha^2 + \alpha^4) \nonumber \\
& + (1 + N) (2 + 4 \alpha^2 + \alpha^4)) (1+\sinh(r)^2) \sinh(r)^4 + (1 + N) (1 + 3 \alpha^2 + \alpha^4) \sinh(r)^6)\nonumber \\
& + \eta^4 (N^2 (\alpha^2 + \alpha^4) (1+\sinh(r)^2)^4 + ((2 + 7 N) \alpha^2 + (4 + 15 N) \alpha^4 \nonumber \\
&+ (1 + 4 N) \alpha^6) (1+\sinh(r)^2)^3 \sinh(r)^2 + (N^3 \alpha^2 + 2 N^2 (1 + N) \alpha^2 \nonumber \\
&+ N^2 (1 + N) (1 + \alpha^2)) (1+\sinh(r)^2)^3 \sinh(r)^2+ (1 + 5 \alpha^2 + 2 \alpha^4 + 2 N^2 (1 + 6 \alpha^2 + 4 \alpha^4)  \nonumber \\
&+ N (3 + 14 \alpha^2 + 4 \alpha^4)) (1+\sinh(r)^2)^3 \sinh(r)^2+ N^2 (1 + N)^2 (1+\sinh(r)^2)^2 \sinh(r)^4  \nonumber \\
&+ (3 + 3 N + 27 \alpha^2 + 30 N \alpha^2 + 24 \alpha^4 + 36 N \alpha^4 + 4 \alpha^6 \nonumber \\
&+ 8 N \alpha^6) (1+\sinh(r)^2)^2 \sinh(r)^4 + (4 \alpha^2 + 14 \alpha^4 + 8 \alpha^6 + \alpha^8) (1+\sinh(r)^2)^2 \sinh(r)^4 \nonumber\\
&+ (1 + 2 N) (3 (1 + \alpha^2)+ N^2 (2 + 4 \alpha^2) + N (5 + 4 \alpha^2)) (1+\sinh(r)^2)^2 \sinh(r)^4 \nonumber \\
& + ((-1 + N) N \alpha^4 + (N^2 + 2 N (1 + N)) (\alpha^2 + \alpha^4) \nonumber \\
&+ (2 N^2 + 6 N (1 + N) + 2 (1 + N)^2) (2 \alpha^2 + \alpha^4) + (2 N (1 + N) + (1 + N)^2) (1 + 3 \alpha^2 + \alpha^4) \nonumber \\
&+ (1 + N) (2 + N) (2 + 4 \alpha^2 + \alpha^4)) (1+\sinh(r)^2)^2 \sinh(r)^4 + (N (1 + N)^2 \alpha^2 + 2 N (1 + N)^2 (1 + \alpha^2) \nonumber \\
&+ (1 + N)^3 (1 + \alpha^2)) (1+\sinh(r)^2) \sinh(r)^6 + ((N^2 + N (1 + N)) (3 \alpha^2 + 2 \alpha^4) + ((1 + N)^2 \nonumber \\
&+ (1 + N) (2 + N)) (3 + 7 \alpha^2 + 2 \alpha^4)) (1+\sinh(r)^2) \sinh(r)^6 + (N (4 \alpha^2 + 5 \alpha^4 + \alpha^6) \nonumber \\
&+ 2 (1 + N) (4 \alpha^2 + 5 \alpha^4 + \alpha^6) + (1 + N) (1 + 7 \alpha^2 + 6 \alpha^4 + \alpha^6)) (1+\sinh(r)^2) \sinh(r)^6 \nonumber \\
&+ (1 + N)^2 (1 + 3 \alpha^2 + \alpha^4) \sinh(r)^8))))^{1/2}.
\label{eq:snr}
\end{align}
\end{widetext}
\end{document}